\documentclass[twocolumn,prl,preprintnumbers,superscriptaddress,showpacs]{revtex4}

\usepackage{graphicx}
\usepackage{amsmath,amsfonts,amssymb,amsxtra}
\usepackage{units}

\makeatletter
\DeclareRobustCommand{\chemical}[1]{%
  {\(\m@th
   \edef\resetfontdimens{\noexpand\)%
       \fontdimen16\textfont2=\the\fontdimen16\textfont2
       \fontdimen17\textfont2=\the\fontdimen17\textfont2\relax}%
   \fontdimen16\textfont2=2.7pt \fontdimen17\textfont2=2.7pt
   \mathrm{#1}%
   \resetfontdimens}}
\DeclareRobustCommand{\bchemical}[1]{%
  {\(\m@th
   \edef\resetfontdimens{\noexpand\)%
       \fontdimen16\textfont2=\the\fontdimen16\textfont2
       \fontdimen17\textfont2=\the\fontdimen17\textfont2\relax}%
   \fontdimen16\textfont2=2.7pt \fontdimen17\textfont2=2.7pt
   \mathbf{#1}%
   \resetfontdimens}}
\makeatother

\newcommand{\vQ}{\chemical{{\bf Q}}}

\begin{document}

\textheight 24.40 true cm

\title{ Control of chiral magnetism in multiferroic MnWO$_4$ through an electric field}

\author{T. Finger}
\author{D. Senff}
\affiliation{%
{II}. Physikalisches Institut, Universit\"at zu K\"oln,
Z\"ulpicher Str. 77, D-50937 K\"oln, Germany}
\author{K. Schmalzl}
\author{ W. Schmidt}
\affiliation{Institut f\"ur Festk\"orperforschung,
Forschungszentrum J\"ulich GmbH, JCNS at ILL, 38042 Grenoble Cedex
9, France}
\author{L.P. Regnault}
\affiliation{Institut Nanosciences et cryog\'enie, SPSMS-MDN,
CEA-Grenoble, DRFMC-SPSMS-MDN, F-38054 Grenoble Cedex 9, France }
\author{P. Becker}
\author{L. Bohat\`y}
\affiliation{Institut f\"ur Kristallographie, Universit\"at zu
K\"oln, Z\"ulpicher Str. 49b, D-50674 K\"oln, Germany}
\author{M. Braden}%
\email{braden@ph2.uni-koeln.de}%
\affiliation{%
{II}. Physikalisches Institut, Universit\"at zu K\"oln,
Z\"ulpicher Str. 77, D-50937 K\"oln, Germany}

\date{\today}
\begin{abstract}

The chiral components in the magnetic order in multiferroic
MnWO$_4$ have been studied by neutron diffraction using spherical
polarization analysis as a function of temperature and of external
electric field. We show that sufficiently close to the
ferroelectric transition it is possible to switch the chiral
component by applying moderate electric fields at constant
temperature. Full hysteresis cycles can be observed which indicate
strong pinning of the magnetic order. MnWO$_4$, furthermore,
exhibits a magnetoelectric memory effect across heating into the
paramagnetic and paraelectric phase.

\end{abstract}

 \maketitle

Magnetoelectric materials allow one to tune the electric
polarization by an external magnetic field and to tune magnetic
polarization by an electric field\cite{1,2,2b}. In particular, the
control of magnetic order by an electric field has a strong
application potential in the context of data storage, but in spite
of strong efforts no suitable materials have been discovered so
far \cite{2,2b}.

Concerning the recently discovered multiferroic transition-metal
oxides \cite{2b,3} it has been well established that the
ferroelectric polarization can be modified by an external magnetic
field \cite{3,7}.
Using polarized neutron scattering in TbMnO$_3$ \cite{7} ,
LiCu$_2$O$_2$ \cite{8} and in MnWO$_4$ \cite{sagayama} it has also
been shown that the chiral component of the magnetic order can be
poled by an electric field when cooling through the ferroelectric
transition. However, in general there have been only very few
reports on a change of a magnetic order induced by applying an
electric field at constant temperature \cite{9,10,10b,11}.
Furthermore, it was shown that by varying the electric field in
MnWO$_4$ one may induce a hysteresis in the ferroelectric
polarization \cite{kundys} and in the second harmonic generation
associated with magnetic domains \cite{meier}. However, the direct
observation of the electric-field induced switching of the chiral
magnetism at constant temperature has not been reported so far in
the spiral multiferroics, although this effect is most important
in view of applications \cite{2b,3}.

\begin{figure}
\includegraphics[width=0.96\columnwidth,angle=0]{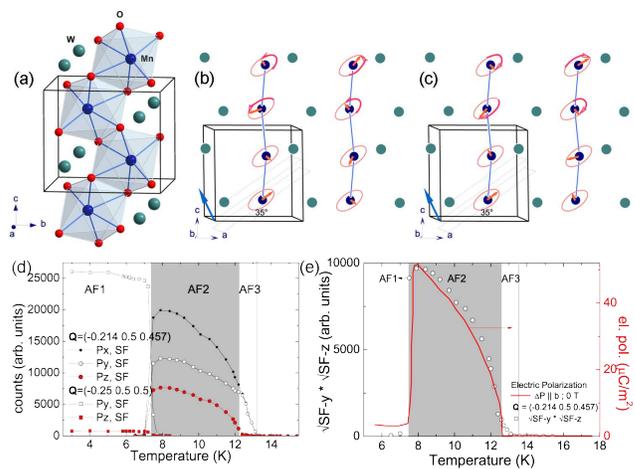}
\vskip -0.1 true cm \caption{(color online) (a) Crystal and (b and
c) magnetic structure in MnWO$_4$; b) and c) show the two opposite
chiral arrangements; d) and e) temperature dependence of magnetic
scattering at $q_{ic}$ and $q_{c}$ studied with polarized neutron
diffraction. The experiment was performed in the
(0,1,0),(-0.214,0,0.457) scattering plane; part d) shows the
magnetic scattering polarized parallel to the easy axis and that
parallel to $b$ and part e) the product of the magnetic components
along these two directions,
$\sqrt{\sigma_{yy}^{\uparrow\downarrow}} \cdot
\sqrt{\sigma_{zz}^{\uparrow\downarrow}}$, which scales well with
the ferroelectric polarization \cite{15}. }
\end{figure}

In most of the recently discovered multiferroics, the
ferroelectric polarization can be explained by the inverse
Dzyaloshinski-Moriya effect \cite{12},
where the induced electric polarization of a single pair of spins
${ S_i},{ S_j}$ separated by a distance vector ${\bf r}_{i,j}$ is
given by \cite{12}:
\begin{equation}\label{eq-katsura}
    {\bf P}_{FE} \propto \bf{r}_{ij} \times (\bf{S}_i \times
    \bf{S}_j).
\end{equation}

The required non-collinear magnetic structure may arise from
strong frustration. Since in addition the interaction (1) is only
a second-order effect, the ferroelectric polarization is rather
small in these materials. In the REMnO$_3$ \cite{3,6} series and
in MnWO$_4$ \cite{14,15,16} the electric polarization is about two
to three orders of magnitude smaller than that in a standard
ferroelectric perovskite such as BaTiO$_3$,
hindering the observation of
electric-field induced effects in the magnetic structure.
Nevertheless, we show in this work that it is possible in these
chiral multiferroics to switch the magnetic order by the
application of a moderate electric field at constant temperature.

The magnetic order in MnWO$_4$ (space group P2/c, a=4.835\AA , b=5.762\AA , c=4.992\AA \ and $\beta$=91.08$^o$) has been determined by neutron diffraction
\cite{18}. Upon cooling, MnWO$_4$ first undergoes a transition into an incommensurate magnetic phase labelled AF3 with propagation vector ${\bf
q}_{ic}$=(-0.214,0.5,0.457) and collinear moments aligned in the $a,c$-plane, T$_{AF3}$=13.2\ K. At the second transition, T$_{AF2}$=12.3\ K, the $b$-component
develops giving rise to a still incommensurate but non-collinear phase, AF2. At the transition AF3$\rightarrow$AF2 the ferroelectric polarization develops, as
independently discovered by three groups \cite{14,15,16}. At further cooling, the magnetic order transforms into a commensurate collinear AF1 state with ${\bf
q}_{c}$=(-0.25,0.5,0.5), T$_{AF1}$=7.0 K .

The polarized neutron-diffraction experiments were performed on
the IN12 cold triple-axis spectrometer at the Institut
Laue-Langevin using either Helmholtz coils or the zero-field
Cryopad for polarization analysis.  An untwinned single crystal
\cite{14}  was set in a $(0,1,0),(-0.214,0,0.457)$ scattering
plane. On a structural Bragg reflection we have determined the
flipping ratio of our polarization setup to 40 and 35 in the two
experimental runs.

The unpolarized neutron-scattering intensity is given by the
square of the magnetic structure factor, $ {\bf M}_\perp ({\bf
Q})=r_0N^{1/2}\sum_{j}{[{\bf M}_j-({\bf Q}\cdot{\bf
M}_j)\cdot{{\bf Q}\over{Q^2}}]e^{i{\bf Q}\cdot{\bf R_j}}}$, where
$r_0=5.4fm$, ${\bf Q}$ denotes the wave vector, and the sum runs
over the atoms in the cell with complex moment ${\bf M}_j$ at
position ${\bf R}_j$. We use the common cartesian coordinate
system with $\bf x$ along ${\bf Q}$, $\bf y$ in the scattering
plane but perpendicular to ${\bf Q}$ and $\bf z$
vertical. 
The polarization analysis \cite{19} adds additional selection
rules :~ 
In the spin-flip (SF) scattering the contributing magnetization
must be perpendicular to the neutron polarization. By measuring
the three SF channels for ${\bf P} || {\bf
x}=$(-0.214,0.5,0.457)$=$\vQ , ${\bf P} || {\bf y}$  and ${\bf P}
|| {\bf z}$, we may thus follow the components as a function of
temperature, see Fig 1. In the AF3 phase the magnetic moment
aligns along an easy axis, $\bf e_{easy}$, in the a,c-plane nearly
perpendicular to the propagation vector ${\bf q}_{ic}$ (the angle
amounts to 83$^o$). Therefore, almost all elastic magnetic
scattering is found in the ${\bf P} || {\bf y}$ channel, and no
magnetic signal is found for ${\bf P} || {\bf z}$. The latter
channel directly senses the $b$-component and becomes finite upon
the phase transition into the AF2 phase. The ferroelectric
polarization \cite{14,15,16} only crudely scales with the
$b$-component, see Fig 1d). At the transition into the AF1 phase
the $b$-component measured in the ${\bf P} || {\bf z}$ channel
disappears as does the ferroelectric polarization \cite{14,15,16}.

The three-dimensional polarization analysis using Cryopad allows
one to determine the full polarization tensor by analyzing the
outgoing and incoming polarization independently \cite{19}.
$\sigma_{ij}^{\downarrow\uparrow}$ denotes the intensity in the
channel with the outgoing polarization along $j$ when the incoming
polarization is set along $i$ with the arrows indicating the
directions of polarizations. The magnetic scattering can be
decomposed into the $M_y ({\bf Q}) \cdot M_y^\ast ({\bf Q})$ and
$M_z ({\bf Q})\cdot M_z^\ast ({\bf Q})$ contributions and the
chiral term, $ {\bf M}_{ch} ({\bf Q})=i\lbrace{\bf M}_{\perp}
({\bf Q})\times{\bf M}_{\perp}^{\ast}({\bf Q})\rbrace $, which
rotates the neutron polarization towards the scattering vector,
and which possesses only a finite $x$ component ${ M}_{ch} ({\bf
Q})$.

The chiral term is determined in three  spherical polarization
channels. In comparison to the total magnetic scattering it is
measured in the $xx$ channel as
$\sigma_{xx}^{\downarrow\uparrow}=\lbrack{M_y ({\bf Q})\cdot
M_y^\ast ({\bf Q})+ {M_z ({\bf Q})}\cdot M_z^\ast ({\bf
Q})}\rbrack -{ M}_{ch}({\bf Q})$ and
$\sigma_{xx}^{\uparrow\downarrow}=\lbrack{M_y ({\bf Q})}\cdot
M_y^\ast ({\bf Q})+ {M_z ({\bf Q})\cdot M_z^\ast ({\bf
Q})}\rbrack+{M}_{ch}({\bf Q})$ by:
\begin{equation}\label{chiral_xx}
r_{chir}={{M}_{ch}({\bf Q})\over { {M_y ({\bf Q})}\cdot M_y^\ast
({\bf Q})+ {M_z ({\bf Q})}\cdot M_z^\ast ({\bf Q}) }}=
{{\sigma_{xx}^{\uparrow\downarrow}-\sigma_{xx}^{\downarrow\uparrow}}
\over
{\sigma_{xx}^{\uparrow\downarrow}+\sigma_{xx}^{\downarrow\uparrow}
}}.
\end{equation}
Alternatively, the chiral contribution is measured at the
non-diagonal components of the polarization tensor :
\begin{equation}\label{chiral_yx}
r_{chir}=
{{\sigma_{yx}^{\uparrow\downarrow}-\sigma_{yx}^{\uparrow\uparrow}}
\over
{\sigma_{yx}^{\uparrow\downarrow}+\sigma_{yx}^{\uparrow\uparrow}
}}=
{{\sigma_{yx}^{\downarrow\uparrow}-\sigma_{yx}^{\downarrow\downarrow}}
\over
{\sigma_{yx}^{\downarrow\uparrow}+\sigma_{yx}^{\downarrow\downarrow}
}},
\end{equation}
in the $yx$ and similarly in the $zx$ channels. In equations (2)
and (3) we may neglect any nuclear contribution as the Bragg-peaks
studied are purely magnetic. Polarized neutron scattering directly
probes the chiral term.

\begin{figure}
~\vskip -0.2 cm
\includegraphics[width=0.999\columnwidth,angle=0]{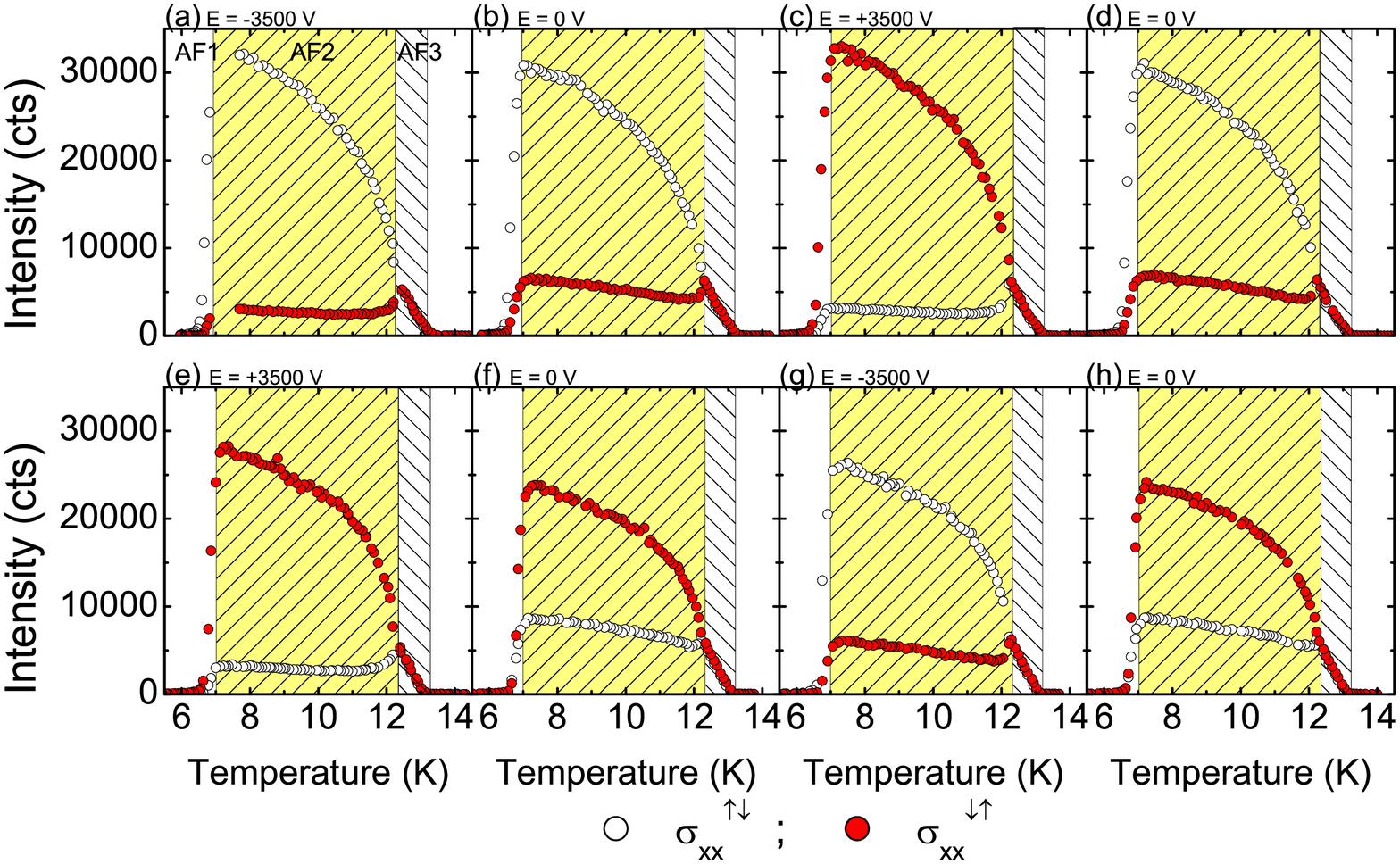}
\caption{(color online)  Temperature dependence of the magnetic
scattering in the ${\sigma_{xx}^{\uparrow\downarrow}}$ and
${\sigma_{xx}^{\downarrow\uparrow}}$ channels measured upon
cooling the MnWO$_4$ crystal with and without electric field. Note
that the intensity in the two $xx$ spin-flip channels are
proportional to the volumes of the two chiral arrangements A) and
B) shown in Fig. 1b) and 1c). After a first cooling in negative
field the thermal cycles a)-d) were successively recorded by
cooling in -3500, 0, +3500 and 0\ V. Before starting the second
run of thermal cycles, e)-h), we heated the sample up to room
temperature and then measured the cooling cycles  in +3500, 0,
-3500, and 0V.}
\end{figure}

If the magnetic order in the sample crystal is a perfect
transverse spiral with the spiral plane perpendicular to the
propagation vector $\bf{q}_{ic}$ parallel to \vQ , the spin flip
scattering $\sigma_{xx}^{\downarrow\uparrow}$ or
$\sigma_{xx}^{\uparrow\downarrow}$ is finite for only one incident
polarization, since ${M}_{ch}({\bf Q})$ and ${ {M_y ({\bf
Q})}\cdot M_y^\ast ({\bf Q})+ {M_z ({\bf Q})}\cdot M_z^\ast ({\bf
Q}) }$ are of the same absolute size yielding $r_{chir}$=$\pm1$
\cite{remark-chiral}. If the geometrical condition that \vQ \ is
perpendicular to the chiral plane, is no longer fulfilled the
chiral contribution to the scattering $r_{chir}$ is reduced even
for the ideal monodomain transverse spiral. In order to detect a
strong chiral term one needs to chose an appropriate scattering
vector. Furthermore, with a real crystal it is necessary to align
the chiral domains, which in the multiferroics can be obtained by
applying an electric field \cite{7,8,sagayama}. After cooling the
sample to T=7.7\ K applying an electric field of 3500V/4mm we have
measured the full three-dimensional polarization matrix at ${\bf
Q}$=(-0.214,0.5,0.457) and (-0.214,1.5,0.457) as well as at the
(0,2,0) structural Bragg reflections. At the first magnetic
reflection the chiral contribution is dominant: the obtained
chiral ratio amounts to 80.8\%. We also find strong contributions
in the ${yx}$ and ${zx}$ channels, which, according to equation
(3) indicate $r_{chir}$=80.3\% and 80.6\%, respectively, in
perfect agreement with the value found by the diagonal term. The
chiral contributions are much weaker at the second reflection
since the large $b$-component of the scattering vector suppresses
the scattering strength of this magnetic component compared to
that along  $\bf e_{easy}$. For \vQ = (-0.214,1.5,0.457), we only
find chiral contributions of 31.0, 32.9 and 32.3\% in the $xx$,
$yx$ and $zx$ channels, respectively. In the refinements of the
magnetic structure \cite{18} it was not possible to determine the
phase between the $\bf e_{easy}$ and $b$ components of the ordered
magnetic moment, which however determines the collinear or chiral
nature of the magnetic structure and thereby the strength of the
multiferroic coupling, see equation (1). Analyzing the full
spherical polarization tensor we may confirm that the phase
between the two components is close to 90$^o$ corresponding to a
chiral arrangement. Therefore, we may confirm that equation (1)
fully explains the occurrence and the direction of the
ferroelectric polarization in MnWO$_4$ \cite{remark-calculation}.
Indeed, the product of the magnetic components along the easy
direction and along $b$ measured by
$\sqrt{\sigma_{yy}^{\uparrow\downarrow}} \cdot
\sqrt{\sigma_{zz}^{\uparrow\downarrow}}$ scales very well with the
temperature dependence of the ferroelectric polarization, see Fig.
1e). From the depolarization of the beam polarized initially along
${y}$ or $z$ we further estimate that with the applied field one
obtains a nearly perfect alignment of the chiral component; only
$\sim$5\% of the sample remain in the opposed chiral state.

\begin{figure}
\includegraphics[width=0.999\columnwidth,angle=0]{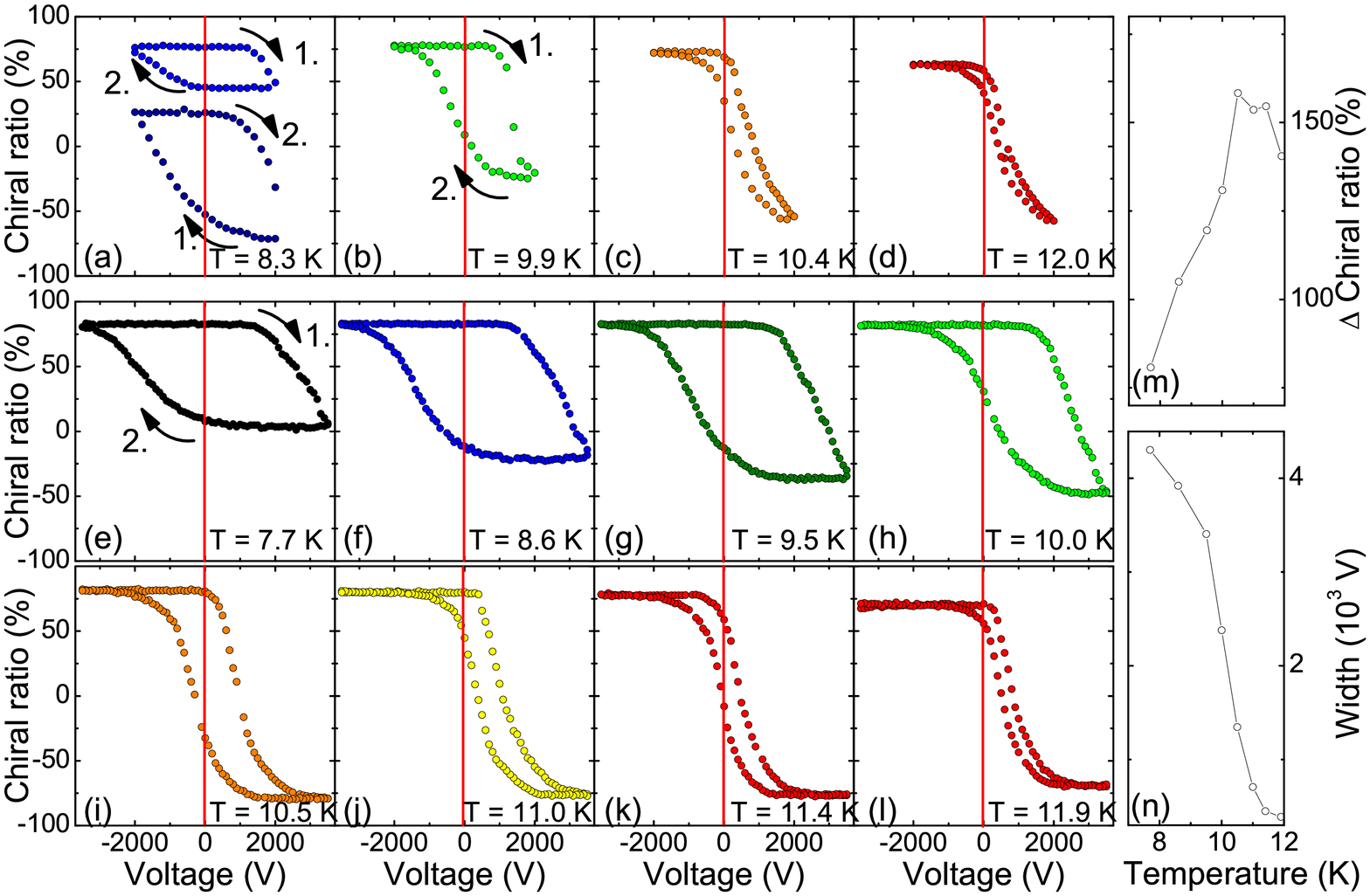}
\caption{(color online)  Hysteresis curves obtained by measuring
the chiral ratio as a function of external electric field at
constant temperature; the hysteresis cycles a)-d) were recorded
after field cooling in E=-2000V (only the lower cycle in a) was
recorded after cooling in +2000V. In a second experiment the
hysteresis curves e)-l) were recorded after cooling in -3500V; m)
and n) present the electric-field induced difference in the chiral
ratio and the width of the hysteresis measured at the middle value
as function of temperature, respectively.}
\end{figure}


In various heating-cooling cycles we found that MnWO$_4$ remembers
its chiral state even after heating into the paramagnetic and
paraelectric phase. The memory and hysteresis behavior of MnWO$_4$
appears to be very complex and is very difficult to be fully
reproduced as it seems to depend on the cooling velocity across
the magnetic transitions. To illustrate the complex memory effect
in MnWO$_4$ we show in Fig. 2 two sets of thermal cycles, Fig. 2
a)-d) and e)-h), which were separated by heating to high
temperature ($\sim$295\ K). First we cooled the crystal in -3500V
obtaining the almost perfectly aligned chiral component which we
label arrangement A), see Fig. 1b) and c). After heating to 15 K,
i.e. into the paramagnetic and paraelectric state, and re-cooling
in zero field we find the identical chiral component, see Fig.
2b). Heating once more and re-cooling in the opposed field allows
one to fully switch the chiral alignment to arrangement B).
Finally another heating to 15\ K with successive zero-field
cooling results in the same arrangement A) although the preceding
+3500V experiment yielded arrangement B). This clearly documents
that the crystal remembers the alignment of the chiral
contribution even in the paraelectric and paramagnetic phase.
Apparently the first cooling in electric field results in a
preference of the crystal for arrangement A) which is not erased
by forcing the sample with the opposed field into arrangement B).
After heating the sample to room-temperature we performed a
similar series of thermal cycling starting with the positive
field. Again one may pole the sample into both arrangements,
although the alignment is less perfect. But now the state B) is
the one observed in the zero-field cooling cycles independently of
the preceding direction of the electric field. One can force the
sample crystal into a preferred chiral arrangement depending on a
first cooling from high temperature (most likely cooling
velocity). This finding resembles the previously reported memory
effect across the paraelectric collinear magnetic phase AF1
\cite{taniguchi2009} and the magnetic-field driven reversibility
\cite{meier}. The fact that the preference is robust against
heating deeply into the paramagnetic state, however, seems to
exclude the given interpretation \cite{taniguchi2009} in terms of
ferroelectric embryos. It appears more likely that the hysteresis
and preference arise from domain-wall pinning by defects, see
below.

In two sets of experiments we studied the possibility to control
the chiral arrangement by varying the electric field at constant
temperature. After cooling the sample in negative voltage we have
measured a hysteresis cycle of the chiral ratio versus electric
field at T=8.3\ K. When fully reducing the voltage to zero the
chiral ratio is unchanged; the voltage may even be inverted and
increased to +1000V without any significant change in the chiral
component documenting the effective pinning of the magnetism in
MnWO$_4$. But further increase of the voltage significantly
reduces the chiral ratio. Upon cycling the voltage back to the
initial negative value the chiral ratio rapidly approaches the
starting value. A quite different hysteresis curve is observed
when the cycle is recorded after cooling with a positive voltage
from about 15\ K, following cooling from room temperature in
negative voltage, see Fig. 3a). The initial chiral term is of
opposite sign but of the same size as that obtained with the
negative voltage, but when lowering the voltage the chiral term
immediately diminishes and even fully changes sign when the
voltage is increased in the negative direction. When driving the
voltage from -2000V back to zero the weaker positive chiral ratio
remains unchanged and only partially recovers the initial value
for increasing the voltage to +2000V. Other hysteresis cycles at
higher temperatures were obtained after field-cooling at -2000V
from 15\ K. When approaching the AF2 to AF3 transition larger
effects are induced in these cycles and the width of the
hysteresis becomes smaller, but all hysteresis curves remain
asymmetric indicating the preferred chiral arrangement. At T=12.0\
K we can induce a complete inversion of the chiral arrangement
through the inversion of the electric field. Such processes form
the basis of the desired application of multiferroics in data
storage techniques.

In a following run we recorded several hysteresis cycles after cooling from high temperature in negative voltage attaining U=-3500V at low temperatures, see Figure
3e)-3l). The same asymmetric hysteresis curves are obtained and again it is possible to fully control the chiral component by the electric field. At T=10.5 K nearly
identical hysteresis cycles were obtained at the four \vQ -positions ($\pm$0.214,0.5,$\mp$0.457)  and ($\pm$0.214,-0.5,$\mp$0.457). With increasing temperature the
height of the hysteresis passes a maximum as the control is facilitated closer to the magnetic transition whereas the size of the chiral component diminishes, see
Fig. 3m). The width of the asymmetric hysteresis continuously decreases upon approaching the paraelectric phase as one may expect due to a weaker pinning, see Fig.
3n).

The hysteresis cycles  shown in Fig. 3a-3l) offer a view on the
processes pinning the magnetism in the multiferroic material. Well
below $T_{AF2}$, the chiral domains seem to be efficiently pinned,
and an inversion cannot be obtained in our large crystal with the
moderate electric fields. Close to the transition the full
magnetic inversion is possible but the hysteresis cycle remains
very asymmetric. Even after passing into the paramagnetic phase
the sample crystal exhibits a pronounced memory for the chiral
domains forced in preceding field-cooling cycles. It must be left
to future experiments to study the exact temperature (apparently
well above 30\ K) which is needed in the heating cycle in order to
erase this memory. The strong pinning and asymmetry of the chiral
domains must be based on strong magnetoelastic coupling. If the
ferroelectric polarization is purely electronic in origin the
associated pinning force should be negligible, whereas a
ferroelectric polarization due to ionic displacements should
posses an intrinsic pinning capability. The pinning of the chiral
may also arise from higher harmonic components which we indeed
observe in MnWO$_4$. The second-order harmonics of the magnetic
modulation studied at \vQ =(-0.428,1.0,0.914) shows magnetic and
sizeable nuclear components. The magnetism in MnWO$_4$ is thus not
only associated with an anharmonic magnetic contribution, but also
with a structural modulation. One may speculate that the latter is
important to understand the pinning of the chiral domains.

In conclusion we have studied the impact of an external electric
field on the magnetic structure in MnWO$_4$. The electric poling
of the chiral terms exhibits a memory effect even when heating
into the paramagnetic and paraelectric phase, which most likely
appears due to pinning of domain walls by defects. Most
importantly, we show that one may control the chiral magnetism by
varying the electric field at constant temperature in the
multiferroic phase. It is possible to observe full multiferroic
hysteresis curves.

This work was supported by the Deutsche Forschungsgemeinschaft in
the Sonderforschungsbereich 608.

\end{document}